\documentclass[conference]{IEEEtran}
\IEEEoverridecommandlockouts
\usepackage{cite}
\usepackage{amsmath,amssymb,amsfonts}
\usepackage{algorithmic}
\usepackage{graphicx}
\usepackage{textcomp}
\usepackage{xcolor}
\usepackage[caption=false,font=footnotesize,labelfont=rm,textfont=sf]{subfig}
\def\BibTeX{{\rm B\kern-.05em{\sc i\kern-.025em b}\kern-.08em
    T\kern-.1667em\lower.7ex\hbox{E}\kern-.125emX}}
\begin{document}

\title{Amplifier-Enhanced Memristive Massive MIMO Linear Detector Circuit: An Ultra-Energy-Efficient and Robust-to-Conductance-Error Design
\thanks{This work was supported by the Beijing Municipal Natural Science Foundation under Grant L242013. \textit{(Corresponding author: Shaoshi Yang)}}
\thanks{J.-H. Bi, S. Yang and P. Zhang are with the School of Information and Communication Engineering, Beijing University of Posts and Telecommunications, Beijing 100876, China (e-mails: \{bijiahui, shaoshi.yang, pzhang\}@bupt.edu.cn).}
\thanks{S. Chen is with the School of Electronics and Computer Science, University of Southampton, Southampton SO17 1BJ, U.K. (e-mail: sqc@ecs.soton.ac.uk).}
}

\author{\IEEEauthorblockN{Jia-Hui Bi, Shaoshi~Yang,~\IEEEmembership{Senior Member,~IEEE}, Ping~Zhang,~\IEEEmembership{Fellow,~IEEE}, Sheng Chen,~\IEEEmembership{Life Fellow,~IEEE}}}

\maketitle

\begin{abstract}
The emerging analog matrix computing technology based on memristive crossbar array (MCA) constitutes a revolutionary new computational paradigm applicable to a wide range of domains. Despite the proven applicability of MCA for massive multiple-input multiple-output (MIMO) detection, existing schemes do not take into account the unique characteristics of massive MIMO channel matrix. This oversight makes their computational accuracy highly sensitive to conductance errors of memristive devices, which is unacceptable for massive MIMO receivers. In this paper, we propose an MCA-based circuit design for massive MIMO zero forcing and minimum mean-square error detectors. Unlike the existing MCA-based detectors, we decompose the channel matrix into the product of small-scale and large-scale fading coefficient matrices, thus employing an MCA-based matrix computing module and amplifier circuits to process the two matrices separately. We present two conductance mapping schemes which are crucial but have been overlooked in all prior studies on MCA-based detector circuits. The proposed detector circuit exhibits significantly superior performance to the conventional MCA-based detector circuit, while only incurring negligible additional power consumption. Our proposed detector circuit maintains its advantage in energy efficiency over traditional digital approach by tens to hundreds of times.
\end{abstract}


\section{Introduction}\label{S1}

Massive multiple-input multiple-output (MIMO) technology, whose core idea is to equip base stations (BSs) with a very large number of antennas to support multiuser transmissions, can significantly improve the network capacity and spectrum efficiency, and it has become a cornerstone technology for contemporary and future wireless communication systems. However, utilizing large number of antennas results in high complexity of detection algorithms, posing a notable challenge to the realization of next-generation massive MIMO receivers that are expected to simultaneously achieve high performance, ultra-low latency and low energy consumption. Many detection algorithms have been proposed to reduce detection latency and energy consumption \cite{Survey_50years}. However, these low-complexity algorithms usually suffer from considerable performance loss, and therefore they do not achieve good trade off between high performance and low latency/low energy consumption. Another effective approach involves accelerating MIMO detection at the hardware level. A variety of application specific integrated circuit (ASIC) designs have been presented for massive MIMO detection \cite{ASIC1,ASIC2}. However, as Moore's law gradually reaches its limits, processors based on the traditional complementary metal oxide semiconductor (CMOS) technology are becoming insufficient to meet the demands of next-generation massive MIMO receivers.

On the other avenue, the emerging memristive devices can be integrated into crossbar arrays for analog matrix computing. Combined with operational amplifiers (OAs), memristive crossbar arrays (MCAs) can perform high-dimensional matrix operations, including matrix-vector multiplication (MVM) \cite{MVM_Exploration}, inverse matrix computation \cite{SunZhong_Inv} and pseudoinverse matrix computation \cite{SunZhong_GeneralizedInv}, rapidly and with low power consumption. Given that massive MIMO detectors primarily involve high-dimensional matrix operations, the MCA enables the realization of next-generation massive MIMO receivers with high performance, ultra-low latency and low energy consumption.

{\color{black}The application of MCA in massive MIMO detection is at a nascent stage. Yuan et al. \cite{Memristor_baseband_processors} employed MCA to accelerate the MVM operations in the minimum mean-square error (MMSE) detection algorithm. However, the application of this work is limited, as it relies on another processor to perform matrix inversion operations. Mannocci et al.  \cite{Memristor_MIMO_Acceleration} introduced an MCA-based ridge regression circuit and applied it to the zero forcing (ZF) and regularized ZF detectors. Zuo et al. \cite{Memristor_Precoder} proposed an MCA-based ZF precoding circuit, consisting of an MVM module and an inversion module, and its core idea can be applied to develop a ZF detector circuit. In \cite{Realizing_InMemory_Baseband}, an MCA-based detector circuit with a structure similar to that in \cite{Memristor_MIMO_Acceleration} was proposed to perform ZF and MMSE detection algorithms. Ren et al. \cite{YiHangRen} employed MCA to accelerate the maximum likelihood detector.} However, the existing MCA-based linear detector circuits \cite{Memristor_baseband_processors,Memristor_MIMO_Acceleration,Realizing_InMemory_Baseband,Memristor_Precoder} did not consider the disparity in large-scale fading coefficients (LSFCs) associated with user terminals (UTs) distributed in different locations of a massive MIMO network. This disparity leads to different elements of the matrices computed in MCA-based circuits following probability distributions with distinct variances, which makes the detection performance susceptible to conductance errors.

To solve this problem, in this paper we propose an MCA-based circuit design for massive MIMO linear detectors. Different from the existing MCA-based detector circuits, our circuit design decomposes the channel matrix into the product of the LSFC matrix and the small-scale fading coefficient (SSFC) matrix, and deals with them separately. We also investigate the impacts of the mapping scheme and conductance errors on detection performance. The proposed detector circuit exhibits significantly superior performance to the conventional MCA-based detector circuit, while only incurring negligible additional power consumption. We also demonstrate significant advantage of the proposed circuit over traditional digital approach in terms of energy efficiency.

\section{System Model and Basic Algorithms}\label{S2}

\subsection{System Model}\label{S2.1}

We consider a massive MIMO system, in which the BS is equipped with $R$ antennas to support $K$ single-antenna UTs with $R>K$. The uplink received signals are given by:
\begin{equation}\label{y=Hs+n_Complex} 
	\tilde{\bf{y}} = \tilde{\bf{H}} \tilde{\bf{s}} + \tilde{\bf{n}}, 
\end{equation}
where $\tilde{\bf{y}} \in \mathbb{C}^{R\times 1}$ is the received signal vector, $\tilde{\bf{s}}\! \in\! \mathbb{C}^{K\times 1}$ is the transmitted signal vector sent by the UTs, $\tilde{\bf{H}}\! \in\! \mathbb{C}^{R\times K}$ is the channel matrix, and $\tilde{\bf{n}}\! \in\! \mathbb{C}{^{R\times 1}}$ is a complex additive white Gaussian noise (AWGN) vector with variance $\sigma_n^2$ per element, i.e., $\tilde{\bf{n}}\! \sim\! {\cal CN}(\mathbf{0}, \sigma_n^2 \mathbf{I})$ with $\mathbf{0}$ and $\mathbf{I}$ denoting the zero vector and the identity matrix of appropriate dimensions, respectively.

Let $\lambda_1, \cdots ,\lambda_K$ be the LSFCs between the $K$ UTs and the BS. The channel matrix $\tilde{\bf{H}}$ can be expressed as:
\begin{equation}\label{eqChMa} 
	\tilde{\bf{H}} = \tilde{\bf{G}} \tilde{\bf{\Lambda}},
\end{equation}
where the diagonal matrix $\tilde{\bf{\Lambda}}=\textrm{diag}\big(\sqrt{\lambda_1}, \cdots ,\sqrt{\lambda_K}\big)$ represents the LSFC matrix and ${\tilde{\bf{G}}} \in \mathbb{C}^{R\times K}$ is the SSFC matrix. We consider the typical Rayleigh fading channel model, which means that the elements of $\tilde{\bf{G}}$ follow the zero-mean Gaussian distribution with variance $\sigma_g^2$ per dimension, namely, 
\begin{equation}\label{eqRaFa} 
  \tilde{g}_{i,j} \sim {\cal CN}\big(0,2\sigma_g^2\big) , \, 1\le i\le R, \, 1\le j\le K .
\end{equation}

The complex-valued system model of \eqref{y=Hs+n_Complex} can be expressed in an equivalent real-valued system model of
\begin{equation}\label{eqRVmimo} 
	\bf{y} = \bf{H} \bf{s} + \bf{n},
\end{equation}
where 
\begin{align*}
  & \bf{y} = \left[ \begin{array}{c}
  \Re \big( \tilde{\bf{y}} \big) \\
  \Im \big( \tilde{\bf{y}} \big)
  \end{array} \right] , ~
  \bf{s} = \left[ \begin{array}{c}
  \Re \big( \tilde{\bf{s}} \big) \\
  \Im ( \tilde{\bf{s}} \big)
  \end{array} \right] , ~
  \bf{n} = \left[ \begin{array}{c}
  \Re \big( \tilde{\bf{n}} \big) \\
  \Im \big( \tilde{\bf{n}} \big)
  \end{array} \right] , \\
  & \bf{H} = \left[ \begin{array}{cc}
  \Re \big( \tilde{\bf{H}} \big) & -\Im \big( \tilde{\bf{H}} \big) \\
  \Im \big( \tilde{\bf{H}} \big) &  \Re \big( \tilde{\bf{H}} \big)
  \end{array} \right],
\end{align*}  
in which $\Re (\cdot )$ and  $\Im (\cdot )$ denote the real and imaginary parts of the corresponding arguments, respectively. In particular, the real-valued channel matrix ${\bf{H}}\in \mathbb{R}^{2R\times 2K}$ is given by
\begin{equation}\label{H=GLamda} 
	\bf{H} = \bf{G} \bf{\Lambda},
\end{equation}
where ${\bf{\Lambda}}=\textrm{diag}\big(\sqrt{\lambda_1}, \cdots ,\sqrt{\lambda_{K}}, \sqrt{\lambda_1}, \cdots ,\sqrt{\lambda_{K}}\big)$ and
\begin{align*}
  \bf{G} =& \left[ \begin{array}{cc}
  \Re \big( \tilde{\bf{G}} \big) & -\Im \big( \tilde{\bf{G}} \big) \\
  \Im \big( \tilde{\bf{G}} \big) &  \Re \big( \tilde{\bf{G}} \big)
\end{array} \right] .
\end{align*}

The task of a massive MIMO detector is to estimate $\bf{s}$ from $\bf{y}$ given the CSI. And the CSI is assumed to be perfectly known in this paper .

\subsection{Basic Detection Algorithms}\label{S2.2}

We consider the following two basic linear detection algorithms.

\subsubsection{ZF Algorithm}

The ZF algorithm estimates signals by:
\begin{equation}\label{ZF_Real} 
	\hat{\bf{s}}_{\textrm{ZF}} = \big({\bf{H}}^{\textrm T} {\bf{H}}\big)^{-1} {\bf{H}}^{\textrm T} \bf{y} ,
\end{equation}
where $(\cdot )^{\textrm T}$ represents the transpose operator, $(\cdot )^{-1}$ represents the inverse operator. Upon substituting \eqref{H=GLamda} into \eqref{ZF_Real} we obtain:
\begin{equation}\label{eqZF} 
  \hat{\bf{s}}_{\textrm{ZF}} = {\bf{\Lambda}}^{-1}\big({\bf{G}}^{\textrm T} {\bf{G}}\big)^{-1} {\bf{G}}^{\textrm T} \bf{y}.
\end{equation}

\subsubsection{MMSE Algorithm}

The MMSE algorithm estimates signals by:
\begin{equation} \label{MMSE_Real} 
	\hat{\bf{s}}_{\textrm{MMSE}} = \big({\bf{H}}^{\textrm T} {\bf{H}} + \rho {\bf{I}} \big)^{-1} {\bf{H}}^{\textrm T} {\bf{y}} ,
\end{equation}
where $\rho=\frac{\sigma_{n}^{2}}{p_s}$ and $p_s$ is the average symbol energy of the transmitted signals. Upon substituting \eqref{H=GLamda} into \eqref{MMSE_Real} we obtain:
\begin{equation}\label{eqMMSE} 
	\hat{\bf{s}}_{\textrm{MMSE}} = {\bf{\Lambda}}^{-1}\big({\bf{G}}^{\textrm T} {\bf{G}} + {\bf{P}}\big)^{-1} {\bf{G}}^{\textrm T} \bf{y} ,
\end{equation}
where ${\bf{P}}=\textrm{diag}\big(\frac{\rho}{\lambda_1}, \frac{\rho}{\lambda_2}, \cdots, \frac{\rho}{\lambda_{K}}, \frac{\rho}{\lambda_1}, \frac{\rho}{\lambda_2}, \cdots, \frac{\rho}{\lambda_{K}}\big)$.

\section{Proposed MCA-Based Circuit Design}
\begin{figure}[tp]
  \vspace*{-1mm}
    \centerline{\includegraphics[width=\linewidth]{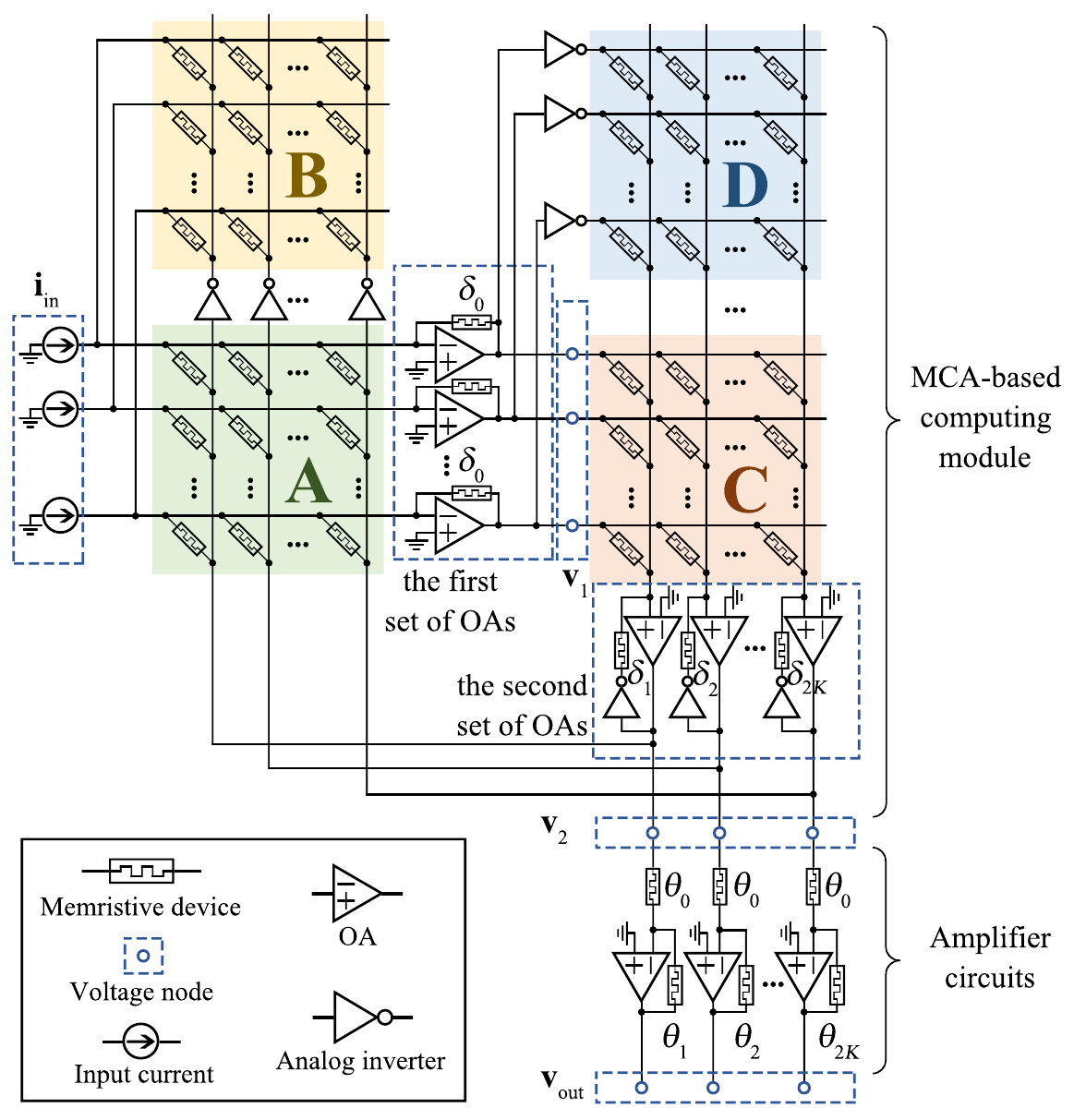}}
  \vspace*{-2mm}
   \caption{The proposed MCA-based detector circuit.}
  \label{detector} 
  \vspace*{-4mm}
\end{figure}

The proposed detector circuit is illustrated in Fig.~\ref{detector}, which is a combination of an MCA-based computing module and $2K$ amplifier circuits. The MCA-based computing module comprises four $2R \times 2K$ MCAs and two sets of OAs, the conductances of the feedback memristive devices of the first set of OAs are all $\delta_0$, and the conductances of the feedback memristive devices of the second set of OAs are $\delta_1, \delta_2, \cdots, \delta_{2K}$. 

Owing to the virtual ground property of OA networks, the voltages at the inverting-input nodes of the first set of OAs and the noninverting-input nodes of the second set of OAs are approximately zeros. Besides, the currents flowing into the inverting-input nodes of the first set of OAs and the noninverting-input nodes of the second set of OAs are approximately zeros owing to the inherent characteristic of OAs. Let $\bf{A}$, $\bf{B}$, $\bf{C}$ and $\bf{D}$ be the conductance matrices of the four MCAs, respectively, ${\bf{v}}_1$ be the output voltages of the first set of OAs, ${\bf{v}}_2$ be the output voltages of the second set of OAs, and ${\bf{i}}_{\textrm{in}}$ be the input currents. Further denote $\bf{E} = \bf{A} - \bf{B}$, $\bf{F} = \bf{C} - \bf{D}$, and ${\bf{\Delta}}_{1} = \textrm{diag}\big(\delta_1, \delta_2, \cdots, \delta_{2K}\big)$. According to Ohm's law and Kirchhoff's law, we have
\begin{equation}\label{PINV_1} 
  {\bf{E}} {\bf{v}}_{2} + {\bf{i}}_{\textrm{in}} + \delta _0 {\bf{v}}_{1} = {\bf{0}} 
\end{equation}
and
\begin{equation}\label{PINV_2} 
  {\bf{F}}^{\textrm T}{\bf{v}}_{1} - {\bf{\Delta}}_{1} {\bf{v}}_{2} = {\bf{0}}.
\end{equation}

Upon substituting \eqref{PINV_2} into \eqref{PINV_1} we obtain:
\begin{equation}\label{PINV_3} 
  {\bf{v}}_{2} = -\big({\bf{F}}^{\textrm T} {\bf{E}} + {\bf{\Delta}}\big)^{-1}{\bf{F}}^{\textrm T} {\bf{i}}_{\textrm{in}},
\end{equation}
where ${\bf{\Delta}}=\textrm{diag}\big(\delta_0\delta_1, \delta_0\delta_2, \cdots, \delta_0\delta_{2K}\big)$.

For the amplifier circuits, let $\theta_1, \theta_2, \cdots, \theta_{2K}$ be the conductances of the feedback memristive devices, and denote $\theta _0$ as the conductance of the memristive devices connected to the output nodes of the second set of OAs. The magnification of the $k$th amplifier circuit is $\frac{\theta_0}{\theta_k}$. The output voltages of the amplifier circuits are:
\begin{equation}\label{PINV_4} 
  {\bf{v}}_{\textrm{out}} = -{\bf{\Theta}}^{-1} {\bf{v}}{_2},
\end{equation}
where ${\bf{\Theta}}=\textrm{diag}\big(\frac{\theta_1}{\theta _0}, \frac{\theta_2}{\theta_0}, \cdots, \frac{\theta_{2K}}{\theta_0}\big)$.

Upon substituting \eqref{PINV_3} into \eqref{PINV_4} we obtain:
\begin{equation}\label{PINV_5} 
  {\bf{v}}{_{\textrm{out}}} = {\bf{\Theta}}^{-1}\big({\bf{F}}^{\textrm T} {\bf{E}} + {\bf{\Delta}} \big)^{-1} {\bf{F}}^{\textrm T} {\bf{i}}_{\textrm{in}}.
\end{equation}

A memristive device is a two-terminal device whose conductance can be changed by charge or flux through it. Using a dedicated program \cite{Program}, the conductance of a memristive device can be set to any desired value within a specified range. By mapping ${\bf{y}}$ onto ${\bf{i}}_{\textrm in}$, mapping ${\bf{G}}$ onto ${\bf{E}}$ and ${\bf{F}}$, setting ${\bf{\Delta}}$ to zeros or mapping ${\bf{P}}$ onto ${\bf{\Delta}}$, and mapping ${\bf{\Lambda}}$ onto ${\bf{\Theta}}$, the result of \eqref{eqZF} or \eqref{eqMMSE} can be obtained by measuring ${\bf{v}}_{\textrm{out}}$. 

The conventional MCA-based detector circuit does not decompose the channel matrix into the product of the LSFC matrix and the SSFC matrix. Obviously, the MCA-based computing module in Fig.~\ref{detector} can be employed as a conventional MCA-based detector circuit to compute \eqref{ZF_Real} or \eqref{MMSE_Real}. Therefore, in the rest of this paper, we employ this module to represent the conventional MCA-based detector circuit for analysis convenience.

\section{Conductance Mapping Schemes}\label{Conductance-Mapping-Scheme} 

The mapped matrix may contain both positive and negative elements, but the device conductance values must remain positive. So we map the matrix onto the difference between two positive conductance matrices, instead of a single conductance matrix. Let the conductance range of memristive devices be $[\omega_{\textrm{min}}, ~ \omega_{\textrm{max}}]$. We define $\omega = \omega_{\textrm{max}} - \omega_{\textrm{min}}$. The scheme for mapping a matrix $\bf{U}$ onto the conductance matrix ${\bf{X}}-{\bf{Z}}$ is:
\begin{equation}\label{eqS4-1} 
  x_{i,j} = \begin{cases}
    {\omega_{\textrm{max}},u_{i,j} > 0} \\
    {\omega_{\textrm{min}},u_{i,j} \leq  0}
    \end{cases}
\end{equation}
and
\begin{equation}\label{eqS4-2} 
  z_{i,j} = x_{i,j} - \alpha u_{i,j},
\end{equation}
where $\alpha$ is the scaling factor. Any conductance that is beyond the conductance range will be clipped to the endpoints.

Process variations and device limitations always lead to conductance errors of memristive devices. The conductance errors can be modeled as Gaussian random variables with mean 0 and variance $\sigma_m^2$ \cite{Error}. Therefore, the impact of conductance errors is equivalent to applying perturbations with variance $\frac{2\sigma_m^2}{\alpha^2}$ to each element of the mapped matrix.

In this section, we propose two conductance mapping schemes, one termed the statistical CSI-based (SCB) scheme, the other termed the instantaneous CSI-based (ICB) scheme.

\subsection{SCB Mapping Scheme}\label{S4.1}

Our SCB scheme selects a fixed scaling factor based on the statistical CSI. Specifically, to map a matrix $\bf{U}$ onto conductance matrices, the SCB scheme calculates the scaling factor by:
\begin{equation}\label{eqS4-3} 
  \alpha = \frac{\omega}{\beta\sigma_u},
\end{equation}
where $\beta$ is the scaling parameter of the SCB scheme and $\sigma_u$ is the standard deviation of the elements of the mapped matrix.

The proposed detector circuit maps $\bf{G}$ onto conductance matrices. For $\bf{G}$, $\sigma_u=\sigma_g$. The conventional detector circuit maps $\bf{H}$ onto conductance matrices. For $\bf{H}$,
\begin{equation}\label{eqS4-5} 
  \sigma_u = \sqrt{\frac{\sum\limits_{k = 1}^{K} {\lambda_k}}{K}} \sigma_g.
\end{equation}

\subsection{ICB Mapping Scheme}\label{S4.2}

Our ICB scheme calculates the scaling factor according to:
\begin{equation}\label{eqS4-7} 
  \alpha=\frac{\omega}{\textrm{max}\{\left|u_{i,j}\right|\}},
\end{equation}
to map $\bf{U}$ onto conductance matrices. Clearly, with this scaling factor, no element of ${\bf U}$ will be clipped. Unlike the SCB scheme, the ICB scheme requires to recalculate the scaling factor once the channel matrix has any changes, thus resulting in greater overhead than the SCB scheme.

\subsection{Discussion}\label{S4.3}

When dealing with a matrix whose elements follow different probability distributions, it becomes challenging to select an appropriate scaling parameter $\beta$ for the SCB mapping scheme. This is because a small scaling parameter leads to a large scaling factor $\alpha$, which is likely to result in a substantial probability of the elements with larger variance being clipped, while a large scaling parameter brings about significant perturbations caused by conductance errors, and the perturbations are particularly severe to the elements with smaller variance. As for the ICB mapping scheme, its scaling parameter is usually decided by the elements with larger variance. Similarly, the perturbations caused by conductance errors are particularly severe to the elements with smaller variance. Evidently, the larger the variance disparity among the different elements of the mapped matrix, the more significant the aforementioned effects become, and the severer the perturbations caused by conductance errors.

For the proposed detector circuit, the elements of the mapped matrix $\bf{G}$ follow the same distribution. In practical scenarios, UTs in a cell always have different distances to the BS, leading to the distinct LSFCs of different UTs. Thus the elements within different columns of the channel matrix $\bf{H}$ follow the probability distributions with different variances. Clearly, the conventional MCA-based detector circuit exhibits a significant variance disparity of the elements of its mapped matrix, which results in severe perturbations caused by conductance errors. This is the reason why we decompose the channel matrix into the product of the LSFC matrix and the SSFC matrix, mapping them separately. It also indicates the superiority of the proposed detector circuit compared to the conventional MCA-based detector circuit.

\section{Simulations}\label{Simulation} 

We consider a multi-user massive MIMO system with 4 UTs and 64 BS antennas, and 64-quadrature-amplitude-modulation (64-QAM) is used in the simulation. The conductance range of memristive devices is $0.1 \rm{\mu}$S $ \sim 30 \rm{\mu}$S. The SPICE simulations in this paper are conducted using LTspice$^\circledR$.

\begin{figure}[bp!]
\vspace*{-7mm}
  \centering
  \subfloat[]{\includegraphics[width=0.5\linewidth]{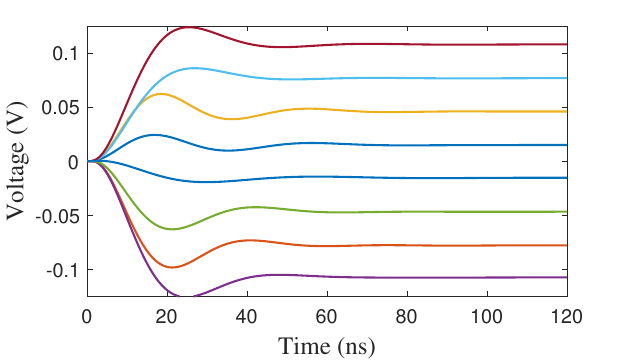}}
  \hfil
  \subfloat[]{\includegraphics[width=0.5\linewidth]{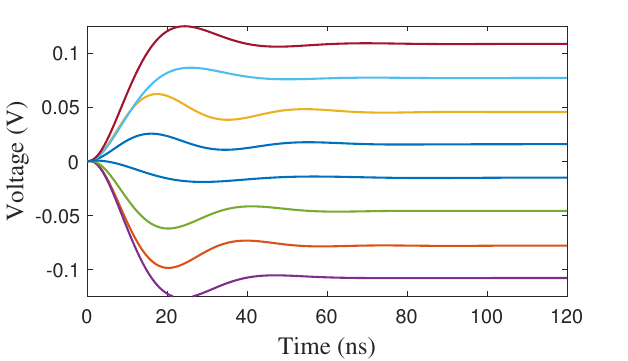}}
\caption{Transient results of output voltages of (a)~the proposed detector circuit, and (b)~the conventional MCA-based detector circuit.}
\label{time} 
\vspace*{-1mm}
\end{figure}

\subsection{Computation Time}

The computation time is an important performance metric for MCA-based detector circuits and we measure it in terms of the convergence time of the circuit. The most critical influencing factor on convergence time is the gain-bandwidth product (GBP) of OAs \cite{Convergence_Time}. The transient results of output voltages of the proposed and conventional MCA-based detector circuits are illustrated in Fig.~\ref{time}, where the OAs have a GBP of 500\,MHz. The proposed circuit exhibits almost identical computation time to that of the conventional MCA-based detector circuit. The computation time of the proposed detector circuit is typically about 80\,ns, and it can be further reduced by increasing the GBP of OAs.

\subsection{Detection Performance}

\begin{figure}[tp]
  \centerline{\includegraphics[width=\linewidth]{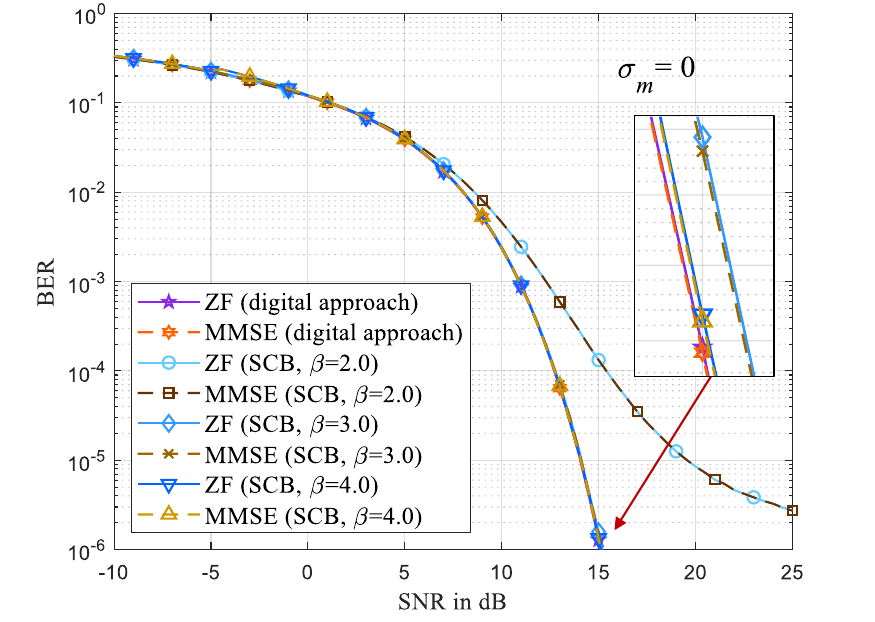}}
  \vspace*{-3mm}
  \caption{BERs of the proposed detector circuit when $\sigma_m=0$ and adopting the SCB scheme.}
  \label{BERvsSNR}
  \vspace*{-3mm}
\end{figure}

\begin{figure}[tp]
  \centerline{\includegraphics[width=\linewidth]{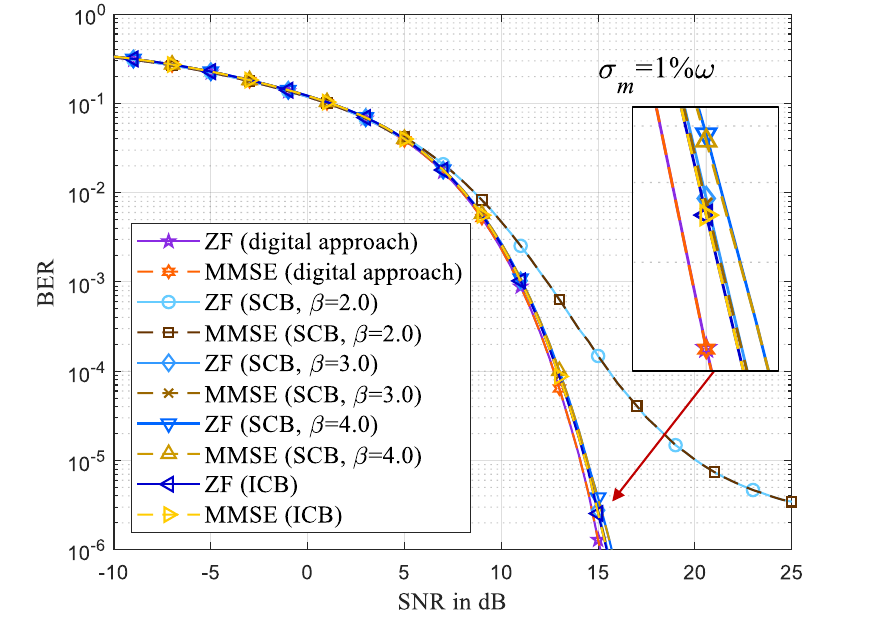}}
  \vspace*{-3mm}
  \caption{BERs of the proposed detector circuit, given $\sigma_m=1$\%$\omega$.}
  \label{BERvsSNR_2}
  \vspace*{-3mm}
\end{figure}

In Fig.~\ref{BERvsSNR}, we compare the bit error rate (BER) performances of the proposed detector circuit adopting the SCB scheme under different scaling parameters, given $\sigma_m=0$, using the digital approach as the benchmark. For the SCB mapping scheme, the larger the scaling parameter, the fewer elements are clipped, and the lower the BER is, i.e., the closer the performance of the proposed detector circuit to digital approach. Specifically, the scaling parameter needs to be at least 3.0 for the proposed detector circuit to ensure satisfactory performance. While the presence of clipped elements increases the BER of the proposed detector circuit, its impact is observable only in high signal-to-noise ratio (SNR). In low SNR, the AWGN remains the primary factor constraining detection performance. Even in the absence of AWGN, clipped elements still cause detection errors, and the BER may exhibit the error floor as the SNR increases.

\begin{figure}[tbp]
  \centerline{\includegraphics[width=\linewidth]{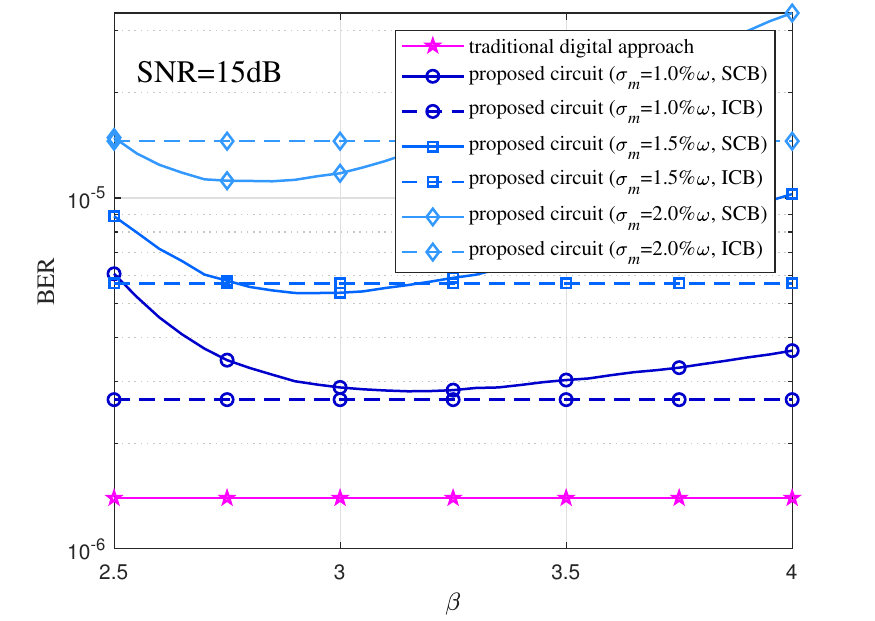}}
  \vspace*{-3mm}
  \caption{BERs of the proposed detector circuit as the functions of the scaling parameter, under different conductance error levels with an SNR of 15\,dB.}
  \label{BERvsBETA}
  \vspace*{-3mm}
\end{figure}

Fig.~\ref{BERvsSNR_2} depicts the BER performances of the proposed detector circuit, given $\sigma_m=1\% \omega$. Simulation results indicate that a larger value of scaling parameter no longer signifies a lower BER. This is due to the fact that a larger scaling parameter implies severer perturbations caused by conductance errors. 

{\color{black}We observe from Fig. \ref{BERvsSNR} and Fig. \ref{BERvsSNR_2} that in the considered scenarios, there is no notable difference in detection performance between ZF and MMSE algorithms. Thus in the remainder of this paper, we use the performance of the MMSE algorithm to represent the performance of the linear detectors considered.}

\begin{figure}[tbp]
  \centerline{\includegraphics[width=\linewidth]{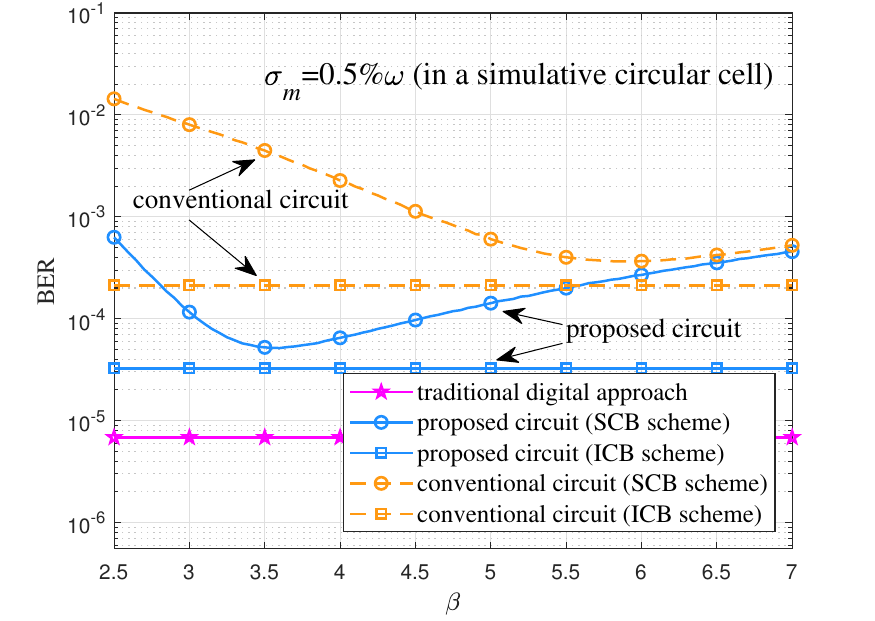}}
  \vspace*{-3mm}
  \caption{BERs of the proposed and conventional detector circuits as the functions of the scaling parameter in a massive MIMO cell, given $\sigma_m=0.5\%\omega$.}
  \label{contrast}
  \vspace*{-3mm}
\end{figure}

To gain further insight into the relationship between the BER and the scaling parameter $\beta$, we exam the BERs as the functions of the scaling parameter for the proposed detector circuit in Fig.~\ref{BERvsBETA}, given different conductance error levels with an SNR of 15\,dB. Simulation results reveal that the BER of the proposed detector circuit adopting the SCB scheme first decreases and then increases as the scaling parameter increases in the presence of conductance errors, because the primary factor constraining detection performance shifts from the clipped elements to the perturbations caused by conductance errors as $\beta$ increases. As expected, the higher the conductance error level, the higher the BER of the detector circuit, regardless whether the SCB scheme or the ICB scheme is adopted. When the conductance error level is low, the BER of the detector circuit adopting the SCB scheme consistently remains higher than that of the ICB scheme. However, when the conductance error level is high, the BER of the detector circuit adopting the ICB scheme is higher than the achievable minimum BER of the SCB scheme.

After investigating the impacts of conductance mapping scheme and conductance errors on detection performance, we demonstrate the performance advantage of the proposed detector circuit over the conventional MCA-based detector circuit. We consider a massive MIMO cell with randomly distributed UTs. The radius of the cell is 150\,m, the uplink carrier frequency is 2\,GHz and the bandwidth is 25\,MHz. The transmitting power of a UT is 20\,dBm. Fig.~\ref{contrast} compares the BER results of the proposed and conventional MCA-based detector circuits as the functions of the scaling parameter, given $\sigma_m=0.5\%\omega$. The results demonstrate that compared to the conventional MCA-based detector circuit, the proposed detector circuit consistently exhibits a significantly lower BER, for both the ICB scheme and SCB scheme.

\subsection{Power Consumption, Computing Performance and Energy Efficiency}

In this experiment, we consider the OA whose static power dissipation is 12\,$\rm{\mu}$W and GBP is 500\,MHz \cite{OAref}. We use current-based digital-to-analog converters (DACs) of \cite{CDACref} to provide input currents for the MCA-based detector circuits. The analog-to-digital converters (ADCs) of \cite{ADC_Array} are used to measure the output voltages. 

\begin{figure}[tbp]
\centerline{\includegraphics[width=\linewidth]{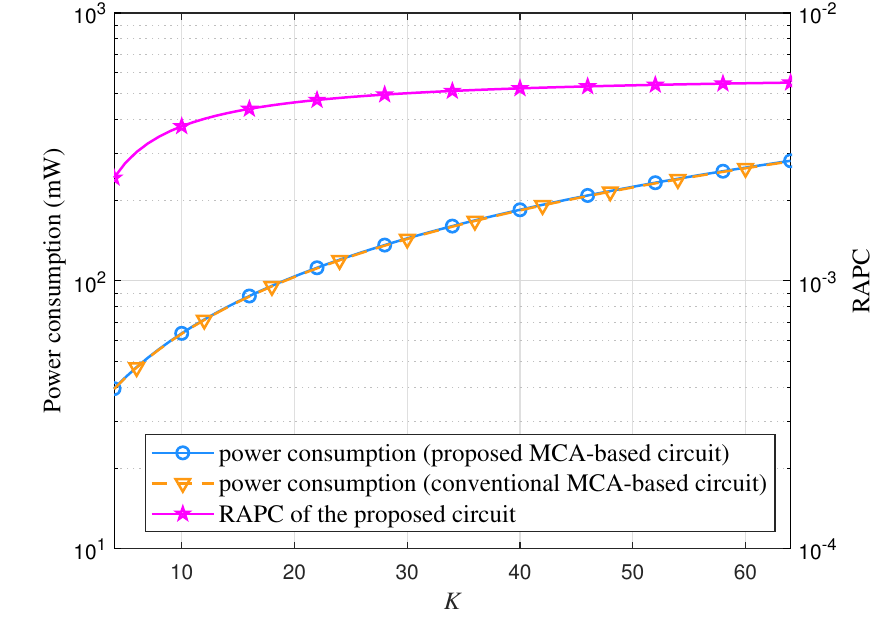}}
\vspace*{-3mm}
\caption{Power consumption results of the proposed and conventional MCA-based detector circuits, as well as the RAPC results of the proposed circuit, as the functions of the number of UTs, $K$.}
 \label{power} 
\vspace*{-3mm}
\end{figure}

The proposed circuit incorporates additional $2K$ amplifier circuits compared to the conventional MCA-based detector circuit. Fig.~\ref{power} depicts the power consumption results of the proposed and conventional MCA-based detector circuits as the functions of the number of UTs, $K$. Meanwhile, Fig.~\ref{power} depicts the relative additional power consumption (RAPC) of the proposed circuit compared to the conventional MCA-based detector circuit. The RAPC of the proposed circuit is less than 0.6\%, which means the additional amplifier circuits of the proposed circuit do not result in significant additional power consumption.

We use the ratio of the number of equivalent floating-point operations (FLOPs) to the computation time of an MCA-based detector circuit as a metric to gauge its computing performance, in which a FLOP is assumed to be either a real multiplication or a real summation. Besides, we use the ratio of the equivalent FLOP number of an MCA-based circuit to the energy consumed during its computation time as a metric to gauge its energy efficiency. The two metrics are measured in tera-FLOPs per second (TOPS) and TOPS/W, respectively. 

\begin{figure}[tbp]
\centerline{\includegraphics[width=\linewidth]{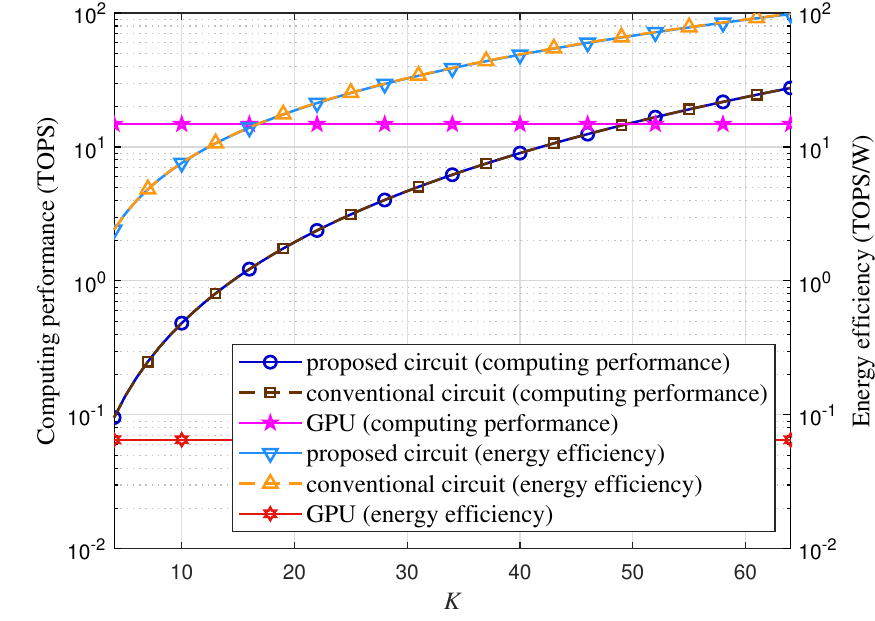}}
\vspace*{-3.5mm}
\caption{Computing performance and energy efficiency results of the proposed and conventional MCA-based detector circuits as the functions of the number of UTs, $K$, using a commercial GPU as the benchmark.}
\label{EE} 
\vspace*{-3mm}
\end{figure}

Fig.~\ref{EE} depicts the computing performance and energy efficiency results of the proposed and conventional MCA-based detector circuits, using the commercial graphic processing unit (GPU) {\it{NVIDIA QUADRO GV100}} \cite{GPU} as the benchmark. There is no significant difference in computing performance and energy efficiency between the proposed and conventional detector circuits. The higher the number of UTs, the higher the dimensions of computed matrices, but the higher the computing performance and the energy efficiency of the MCA-based detector circuits. The MCA-based detector circuits exhibit computing performance advantages over the commercial GPU only when $K$ is relatively large, but their energy efficiency surpasses the GPU by several orders of magnitude.

\section{Conclusions}\label{Conclusion}

We have proposed an MCA-based circuit design for massive MIMO ZF and MMSE detectors. The proposed detector circuit employs an MCA-based matrix computing module and OA-based amplifier circuits to separately deal with the SSFC matrix and the LSFC matrix, significantly reducing the perturbations caused by conductance errors. We have presented two conductance mapping schemes for the MCA-based detector circuits, one termed the SCB scheme and the other termed the ICB scheme. We have investigated the impacts of mapping scheme and conductance errors on detection performance of the proposed detector circuit and have demonstrated the significant performance advantage of our proposed detector circuit over the conventional MCA-based detector circuit. Although the proposed circuit incorporates additional amplifier circuits compared to the conventional MCA-based detector circuit, the additional amplifier circuits do not result in observable additional power consumption. The energy efficiency of the proposed circuit is tens to hundreds of times that of the commercial GPU {\it{NVIDIA QUADRO GV100}}. The core idea of the proposed scheme can also be applied to techniques such as precoding and channel estimation, thus enhancing the applicability of MCA in massive MIMO systems.

\end{document}